\documentclass[12pt]{article}

\catcode`\@=11

\global\arraycolsep=2pt
\usepackage{amsbsy,amssymb,latexsym,amsfonts,amsmath}
\usepackage{graphicx,color}

\allowdisplaybreaks

\begin{document}
\begin{flushright}
\parbox{4.2cm}
{YITP-24-31}
\end{flushright}

\vspace*{0.7cm}

\begin{center}
{ \Large Generalized Free Conformal Field Theories at Infinite Temperature -- In memory of Iwasaki-sensei}
\vspace*{1.5cm}\\
{Yu Nakayama}
\end{center}
\vspace*{1.0cm}
\begin{center}

Department of Physics, Rikkyo University, Toshima, Tokyo 171-8501, Japan

Yukawa Institute for Theoretical Physics,
Kyoto University, Kitashirakawa Oiwakecho, Sakyo-ku, Kyoto 606-8502, Japan

\vspace{3.8cm}
\end{center}

\begin{abstract}
Can space-time symmetries such as Lorentz, dilatation, or conformal symmetry 
 be recovered at infinite temperature? To address this question, we study correlation functions of generalized free conformal field theories (a.k.a free holographic theories in thermal AdS space-time) at infinite temperature. We show that they are broken at the leading order in bosonic correlation functions, but a non-trivial scaling symmetry may emerge.
\end{abstract}

\thispagestyle{empty} 

\setcounter{page}{0}

\newpage

\section{Introduction}
Around a year ago, we received the tragic news that Iwasaki-sensei passed away. During the Covid-19 period, I did not have an opportunity to interact with him, so it was sudden news. Several days later, at dawn, Iwasaki-sensei somehow appeared in my dream suggesting I should study the infinite temperature phase of QCD. We expect that all the symmetries may be recovered at the infinite temperature in finite systems, but is it really the case in QCD? This was his question in the dream. 

I had the privilege to be one of Iwasaki-sensei's last collaborators, studying the relation between the conformal phase of (multi-flavor) massless QCD and the high-temperature phase of (small flavor) massless QCD. Iwasaki-sensei had never mentioned the infinite temperature, though. Obviously, it was {\it my} question unconsciously raised while half-asleep. The immediate answer was that in QCD no propagating degrees of freedom remain in the infinite temperature limit (i.e. fermions and gauge fields would not propagate) and symmetries are restored in a trivial fashion. But I knew that this answer would not satisfy him. Iwasaki-sensei had always looked for a promising direction.

I just recalled that during the collaboration with Iwasaki-sensei on the conformal phase of multi-flavor QCD, as a toy model to demonstrate his idea, I had studied generalized free conformal field theories or free fields in AdS space-time with various IR cut-offs and finite temperature effects. Maybe there is room for something to happen in the infinite temperature limit in certain conformal field theories even though it may not correspond to QCD. In fact, we know in some integrable systems, non-trivial scaling symmetry emerges in the infinite temperature limit. So I wrote a note.

I have had this small note on my desk for some time.  In March 2024, Kanaya-san organized a gathering in memory of Iwasaki-sensei. Kanaya-san said the original plan to organize a workshop failed because Iwasaki-sensei had too many descendants. The seeds Iwasaki-sensei had planted all blossomed beautifully.
After the gathering, returning to Kyoto on the Shinkansen, I suddenly felt like adding a bit to the note. This is it.

\section{Generalized free conformal field theories at infinite temperature}
In finite systems, we can formally argue that all the symmetries make the expectation values invariant in the infinite temperature limit $\beta \to 0$. For a unitary symmetry $U$, we have
\begin{align}
\langle U^\dagger O U \rangle_{\beta =0} = \lim_{\beta \to 0} \mathrm{Tr} U^\dagger O U \frac{e^{-\beta H}}{Z(\beta)}   = Z(0)^{-1}\mathrm{Tr}O =  \langle O \rangle_{\beta =0} \ .
\end{align}
Note that we do not have to take the infinite temperature limit (at least formerly) if the symmetry commutes with the Hamiltonian: we can just use the cyclicity of the trace to show its invariance. 
It was the basis of our intuition that the global symmetry (i.e. chiral symmetry in QCD) will get restored at finite temperature. The point is even if the symmetry does not commute with the Hamiltonian (e.g. Lorentz symmetry or dilatation symmetry), the infinite temperature limit makes the expectation values invariant.

It is, however, not immediately obvious if this intuition applies to space-time symmetries such as Lorentz symmetry or conformal symmetry in quantum field theories with infinite degrees of freedom. Indeed, if we try to evaluate the above formal manipulation line by line in quantum field theories, we encounter infrared divergences. The total Hilbert space of quantum field theories is too large to take the trace. 

Interestingly it is known that, with careful limits taken, some interesting structures may emerge. One is the appearance of the emergent KPZ scaling behavior in integrable spin chains in the infinite temperature limit \cite{DM}. Another is the appearance of "tomperature" in the infinite temperature limit of the double-scaled SYK models \cite{Lin:2022nss}. Furthermore, the appearance of the spontaneous breaking of symmetries in the infinite temperature limit (with potential Nambu-Goldstone phases!) was proposed in \cite{Chai:2020onq}. These examples suggest nontrivial things can happen.

In this section, we study the infinite temperature limit of the generalized free conformal field theories (or free theories in thermal AdS space-time). Let us start with the scalar two-point functions of 
generalized free conformal field theories with conformal dimension $\Delta$ at finite temperature $\beta^{-1}$:
\begin{align}
\langle \Phi(t,\vec{x})\Phi(0,0) \rangle_\beta = \sum_{m \in \mathbb{Z}} \frac{1}{((it- m\beta)^2 + \vec{x}^2)^\Delta}  \ . 
\end{align}
Here $t$ is a real time rather than an imaginary time. This was computed by a ``method of image" with periodic boundary conditions in imaginary time. 

In the infinite temperature limit $\beta \to 0$, we can approximate the sum over $m$ by an integral: more explicitly we evaluate it as 
\begin{align}
    \int \frac{dm}{\beta} \frac{1}{((m-it)^2 + \vec{x}^2)^\Delta} \sim \frac{1}{\beta t^{2\Delta- 1}} F(\Delta-\frac{1}{2},\Delta; \Delta + \frac{1}{2}; \frac{\vec{x}^2}{t^2} )  
\end{align}
when $\vec{x}^2 < t^2$. (If $\vec{x}^2>t^2$, the integral is independent of $t$ and it is proportional to $\beta^{-1} (\vec{x}^2)^{\frac{1}{2}-\Delta}$.)
Note that it is divergent in the $\beta \to 0$ limit, and we have kept the most singular term. 

The two-point function is not Lorentz invariant nor conformal invariant, but it has an interesting emergent scaling behavior. For zero spatial separation, it behaves as $\beta^{-1} t^{-2\Delta + 1}$. In the "light cone" limit $\vec{x}^2 \to t^2 $, it behaves as $(1-\frac{\vec{x}^2}{t^2})^{1-\Delta}$. Thus while it is not Lorentz invariant, it still retains some notion of ``light cone". (One can always try to associate the hypergeometric function with $SL(2,\mathbb{R})$ or hidden conformal symmetry.) The infinite temperature limit is opposite to the OPE limit e.g. studied in \cite{Iliesiu:2018fao}, so even the UV scaling behavior changes.

If we consider a fermionic operator, we have to impose the anti-periodic conditions in imaginary time. Suppressing the spin indices, we obtain the thermal two-point functions of fermionic operators in generalized free field theory at finite temperature as
\begin{align}
\langle \Psi(t,\vec{x})\Psi(0,0) \rangle_\beta = \sum_{m \in \mathbb{Z}}  (-1)^m\frac{1}{((it- m\beta)^2 + \vec{x}^2)^\Delta} \ .
\end{align}
Since it is an alternating sum of an even function, in the $\beta \to 0 $ limit, it is of order  $\frac{\beta t}{(\vec{x}^2 -t^2)^{\Delta +1}} \sim \beta t^{-2\Delta - 1} $ rather than $t^{-2\Delta}$ or $\beta^{-1} t^{-2\Delta +1}$ (for zero spatial separation). The bosonic two-point function was singular but the fermionic two-point function rather vanishes. It is not Lorentz invariant nor conformal invariant, but there is a singular behavior along the ``light cone". 

Let us make a comment on the finite temperature AdS/CFT correspondence. 
Our thermal two-point functions of generalized free conformal field theories correspond to the holographic two-point functions of a free field in thermal AdS space-time. If we, instead, consider the two-point functions in the AdS black brane, the two-point function vanishes exponentially in the infinite temperature limit. See also some recent works on holographic computations on thermal two-point functions \cite{Rodriguez-Gomez:2021pfh}\cite{Rodriguez-Gomez:2021mkk}\cite{Karlsson:2022osn}\cite{Dodelson:2022yvn}\cite{Bhatta:2022wga}\cite{Bajc:2022wws}.\footnote{The thermal AdS and the AdS black hole can be distinguished more transparently in the global space-time as the Hawking-Page transition. Here, on the Poincar\'e patch, the phase depends on a more subtle order of limit.}
This is closer to our intuition that in the infinite temperature limit, all the correlation functions vanish in generic interacting field theories. The fermions do not propagate due to the Matsubara gap with anti-periodic boundary conditions and the gauge fields do not propagate due to the plasma screening. Then the bosons will not be critical because we cannot fine-tune any parameters in the thermal potential.
 If one can tune the theory itself by changing external parameters, we may circumvent the genericity argument, which should be the case in the global symmetry restoration scenario in the infinite temperature limit of quantum field theories.

 Finally, let us mention the potential applications of generalized free conformal field theories in real physics. They most likely appear in non-local systems. The thermal two-point functions of generalized free conformal field theories are not governed by the Virasoro symmetry in two dimensions even though the zero-temperature two-point functions look conformal invariant. The absence of the Virasoro symmetry is related to the absence of the local energy-momentum tensor.
 The long-range statistical models such as the long-range Ising model may realize a structure similar to the one discussed in this note. It may be interesting to numerically study the infinite temperature quantum systems with long-range interaction.

\section{In memory of Iwasaki-sensei}
Over a decade ago, when I was a postdoc at Kavli IPMU, I got a phone call from a secretary, saying ``Listen. You will hear the voice of the former president of Tsukuba University, and the current executive secretary of KEK, Prof. Yoichi Iwasaki. He has something he wants to ask you. Be prepared and answer them politely. Now let me connect the line."
This was the beginning. Naturally, I got nervous\footnote{The phone call reminded me of the 2-26 incident in Japan. Off-guarded by the riot, the emperor himself attempted to call the nearby police office for help, but the officer on duty thought it was a prank call and hung up. Five minutes later, another phone call rang. This time it was from the servant: ``Listen. You will hear the voice of the noblest one in this empire. Be prepared. Now let me connect the line." The officer naturally went pale as I did!} but my fear was unfounded. Iwasaki-sensei just wanted to discuss physics with me. He was so enthusiastic about the conformal phases of multi-flavor QCD that he never stopped explaining his ideas to me. I share the same passion for CFTs in higher dimensions and we immediately hit it off. Later, I learned that Prof. Yanagida had recommended me to Iwasaki-sensei.

While I didn't  recognize it at the time, everyone else thought it was the return of Phoenix. Iwasaki-sensei was famously known as the father of the lattice QCD in Japan owing to his unparalleled success with CP-PACS. Serving important administration roles, not to mention five years of presidency at Tsukuba University, everyone thought he was now long retired from physics research. Not at all. He always had a desire to do physics. 

We have published many works since then \cite{Ishikawa:2013wf}\cite{Ishikawa:2013iia}\cite{Ishikawa:2013tua}\cite{Ishikawa:2015iwa}\cite{Ishikawa:2015nox}\cite{Ishikawa:2017hka}\cite{Ishikawa:2017nwl}, but it was not without trouble. In 2014, Iwasaki-sensei got a heavy injury. Along with his chronic illness he had suffered, the doctor suggested he would never be able to get up from the bed: he should move to alleviate his illness but he could not move because of the injury. With dedicated support from his family and his strong willpower, however, he came back like a Phoenix. He got up from the bed, walked again, wrote emails, ran jobs on the supercomputer, and continued research in physics. The doctor said it was a miracle. 
After his magical recovery, I had an opportunity to be invited to his house in Setagaya, Tokyo. I was warmly hosted with a feast beautifully prepared by Shizue-san. 
They treated me as if I were their child, serving dishes that were all their sons' favorites, including small hamburgers.  Iwasaki-sensei looked as happy as ever.  

At the gathering held in memory of Iwasaki-sensei, one year after his passing away, I learned that for many of us, the last conversation with him was about the research e.g. ``My job is not running smoothly: is there anything we could do?". For me, it was about the comment made by Prof. Sinya Aoki about my latest project with Iwasaki-sensei. Iwasaki-sensei had an idea that near the ``conformal point" such as the zero-temperature limit of the multi-flavor massless QCD or near the chiral phase transition point of the two-flavor massless QCD\footnote{For him, it was ``obvious" (and ``proved numerically by himself \cite{JLQCD:1998qth}") that the chiral phase transition of the two-flavor massless QCD was second order, but I think this is still a matter of debate.}, the scaling behavior of the correlation functions (on the finite lattice) must be better behaved than in non-conformal situations.\footnote{In this scaling analysis, we do {\it not} change the gauge coupling constant, so it is different from the UV continuum limit. Rather it is an IR thermodynamic limit (because physics does not change without renormalization of the UV parameters). The distinction between the two notions may address the concern that Prof. Aoki had.} He thought he could diagnose if a theory is in the conformal phase or not by this scaling analysis. He showed me many plots to support his ideas. He always loved plots: plots are the origin of his inspiration.

I admire Iwasaki-sensei as a theoretical physicist.

\section*{Acknowledgements}
This work is in part supported by JSPS KAKENHI Grant Number 21K03581.

\end{document}